\documentstyle[12pt,epsfig]{article}
\newcommand{\beq}{\begin{equation}}
\newcommand{\eeq}{\end{equation}}
\newcommand{\beqar}{\begin{eqnarray}}
\newcommand{\eeqar}{\end{eqnarray}}
\newcommand{\noi}{\noindent}
\begin{document}
\vspace*{1cm}

\begin{center}
  {\LARGE A Note on CP Invariance and \\ \vspace{3mm}

 Rephasing Invariance\footnote{To appear in the proceedings of 
             ``KEK meeting on CP violation and its origin", 
             (ed. K.~Hagiwara, KEK, 1997)
              } 
}
\end{center}
\vspace*{0.1cm}
\begin{center}
{\large T.~Kurimoto}\footnote{e-mail: krmt@sci.toyama-u.ac.jp}\\
Department of Physics, Faculty of Science,\\
Toyama University,\\
Toyama 930, Japan\\

\vspace*{3cm}
\Large{\bf Abstract} 
\end{center}

CP invariance condition and rephasing invariance in KM scheme 
of CP violation is discussed.   
The CP violation measure in B physics, corresponding to the three angles of unitarity 
triangle are given in rephasing invariant form. With the obtained results it is shown 
that the sum of the three CP angles to be measured at B factories becomes $\pi$  
even if the Kobayashi-Maskwa matrix is not a $3\times 3$ unitary matrix under reasonable 
assumptions.   

\newpage
\section{CP invariance}
It is often said that CP invariance holds if all the coupling constants 
in the theory are real. This is true, but simultaneously we can make some of 
coupling constants complex by rephasing complex fields 
without changing physics. Therefore more accurate definition of 
CP invariance is necessary in particular when we deal with the rephasing 
invariance of physical quantities. In this note we discuss the 
CP violation scheme of Kobayashi-Maskawa (KM)\cite{KM} by taking the freedom 
of rephasing quark fields into account, and give  rephasing invariant 
expressions of the CP violation measures at B meson system, so called the 
CP angles of the unitarity triangle\cite{unit}, in 
the standard model and its extensions.

CP transformation of a quark field $q$ is defined as
\beq
 (CP)q(CP)^{-1} = e^{i\tilde q} \gamma^0 q^C = 
                  e^{i\tilde q}\gamma^0 C (\bar q)^T ,\label{cp}
\eeq 
where $C$ is the charge conjugation matrix and $ e^{i\tilde q}$ is a 
phase factor which in general can be taken arbitrarily and 
depend on the field. The interaction among quark charged currents and 
$W$ boson is given in quark mass eigenstates as
\beq
{\cal L}_W = \frac{g}{\sqrt{2}}\left[
\overline {u_{Li}}\gamma^\mu (V)_{ij}d_{Lj}W^+_\mu 
  + \overline {d_{Lj}}\gamma^\mu (V^*)_{ij}u_{Li}W^-_\mu \right],
\eeq
where $i,j$ are the flavor indices and $V$ is KM matrix.
By the CP transformation given in eq.(\ref{cp}) and 
$(CP)W_\mu^{\pm}(CP)^{-1} = - W^{\mp\mu}$ we have 
\beq
(CP){\cal L}_W (CP)^{-1} = \frac{g}{\sqrt{2}}\left[
\overline {d_{Lj}}\gamma_\mu 
           (V)_{ij}e^{i(\tilde{d}_{Lj}-\tilde{u}_{Li})}
          u_{Li}W^{-\mu}
 + \mbox{ (h.c.)} 
 \right].
\label{wint}
\eeq
We take $\tilde q_L = \tilde q_R \equiv \tilde q$ so that the mass term 
might be CP invariant, then CP invariance requires that the following 
conditions should hold
\beq
(V)_{ij} = (V^*)_{ij}e^{-i(\tilde{d}_j-\tilde{u}_i)},
\label{cpkm}
\eeq
by suitably choosing the CP phases $\tilde q$. If we take 
$\tilde{u}_i=\tilde{d}_j$ for any quark flavor, the  condition of 
CP invariance becomes that $V_{KM}$ should be real, the usual one. 
On the contrary CP invariance means $V$ is pure imaginary 
if we take $\tilde{u}_i=\tilde{d}_j+\pi$. CP violation occurs 
when any choice of $\tilde q$ cannot satisfy 
the condition (\ref{cpkm}). Say in other words, if we can find a set of 
$\tilde q$ which satisfy the condition (\ref{cpkm}), we can make 
KM matrix real by the redefinition of the phases of quarks and the theory 
is CP invariant in the usual manner, 
which shall be explained in the next section.
\section{Rephasing}
Let us consider redefinition of the phases of quarks;
\beq
     q' = e^{-i\hat q} q,
\eeq
where the rephasing angle $\hat q$ is arbitrary and depends on the 
filed. Note that $\hat q$  is independent of the CP phase $\tilde q$ 
discussed before. Physics does not change under this rephasing. 
The Lagrangian (\ref{wint}) becomes by rephasing as 
follows;
\beq
{\cal L}_W = \frac{g}{\sqrt{2}}\left[
\overline {u_{Li}'}\gamma^\mu (V')_{ij}d_{Lj}'W^+_\mu
+\mbox{ (h.c.)} \right],
\eeq
where the transformed KM matrix element is given as  
$(V')_{ij}= e^{i(\hat d_j-\hat u_i)}(V)_{ij}$. Suppose 
the original Lagrangian (\ref{wint}) is CP invariant, that is 
KM matrix $V$ satisfies the condition (\ref{cpkm}) for a set of CP 
phase $\tilde q$, then 
\beq
(V')_{ij} 
               = e^{i(\hat d_j-\hat u_i)} 
                 (V^*)_{ij}e^{-i(\tilde{d}_j-\tilde{u}_i)}
               =  ({V'}^*)_{ij}
                   e^{-i\{ (\tilde{d}_j - 2 \hat d_j) - 
                           (\tilde{u}_i - 2 \hat u_i)
                         \} }         .
\eeq
So we should change the CP phase $\tilde q$ to  $\tilde q-2\hat q$ 
under the rephasing of quarks. If the CP invariant condition 
(\ref{cpkm}) is satisfied,  KM matrix can be made real by 
the rephasing of the quark fields with $\hat q = \tilde q/2$.  
\section{Rephasing invariant expressions of CP angles}
In the case of 3 generations KM matrix is a $3\times3$ unitary matrix so that 
the following condition holds :
\beq
{V_{ub}}^* V_{ud} + {V_{cb}}^* V_{cd} + {V_{tb}}^* V_{td} =0.
\eeq  
We have so-called unitarity triangle by expressing the above condition 
in complex plane.
\begin{figure}[hbtp]
  \begin{center}
    \leavevmode
    \epsfig{figure=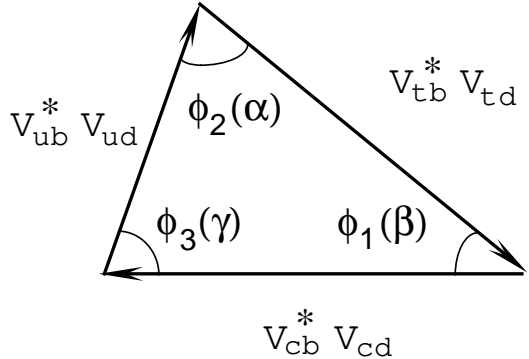, width=7cm}
    \caption{Unitarity triangle}
    \label{fig:unitri}
  \end{center}
\end{figure}
The parameters in KM matrix can be determined 
through the measurements of the sides and the angles of this triangle.
We can test the accuracy of the 3 generation standard model by over-checking 
the consistency of the triangle. If the test fails, we can explore 
new physics beyond the standard model from the inconsistency\cite{new}. 
This is one of the main aims of the $B$ factory projects at KEK\cite{KEKB} 
and SLAC\cite{SLACB}. 

The angles of the unitarity triangle are geometrically defined as
\begin{eqnarray}
  \label{cpangles}
  \phi_1 \ (\beta) &\equiv &  \mbox{arg}[{V_{cb}}^* V_{cd}]
                                - \mbox{arg}[{V_{tb}}^*V_{td}] - \pi, \\
  \phi_2 \ (\alpha) &\equiv &  \mbox{arg}[{V_{tb}}^* V_{td}]
                                - \mbox{arg}[{V_{ub}}^*V_{ud}] + \pi, \\
  \phi_3 \ (\gamma) &\equiv &  \mbox{arg}[{V_{ub}}^* V_{ud}]
                                - \mbox{arg}[{V_{cb}}^*V_{cd}] + \pi .
\end{eqnarray}
These angles do not necessarily agree with the CP angles to be measured 
in experiments, which we call $\tilde\phi_i$ $(i=1,2,3)$ below, if a new physics 
contributes significantly to CP violation.  
New physics can affect the CP angles through 
(i) $B^0$-$\overline {B^0}$ mixing as in the case of SUSY standard 
models\cite{bbss} , (ii) $b$ decay as in $SU(2)_L \times SU(2)_R \times U(1)$ 
models\cite{lrmd} or (iii) deformation of KM matrix as in 4 generation 
models\cite{fgen} or extra vector-like quark models\cite{vecq}.
Let us see the differences between $\phi_i$ and $\tilde\phi_i$ $(i=1,2,3)$ 
under the following assumptions;

\vspace*{3mm}\noi
{\it Quark decay amplitude has the same 
phase as those given by the corresponding tree level $W$ boson exchange diagram 
up to minor corrections except for the $\Delta I=1/2$ penguin type contribution.}

\vspace*{2mm}\noi
{\it D meson decay is dominated by the tree level $W$ boson interaction with 
negligible CP violation. 
}

\vspace*{3mm}\noi
Without the first assumption the CP angles will become irrelevant to KM matrix elements
and the details of new physics should be necessary for discussion. The second one is necessary 
because we use D meson decay to obtain $\tilde \phi_3$ at B factories.
\subsection{$\phi_1 \ (\beta)$}
The CP angle $\tilde\phi_1$ corresponding to $\phi_1$ $(\beta)$ can be 
 measured through 
time dependent $CP$ asymmetry of the neutral $B$ meson decay into a $CP$ 
eigenstate, $J/\Psi K_S$ \cite{cpang} :
\begin{eqnarray}
 Asy[J/\Psi K_S] &\equiv& 
\frac{\Gamma[B^0 (t) \rightarrow J/\Psi K_S] - 
      \Gamma[\overline{B^0}  (t)\rightarrow J/\Psi K_S]
      }{
      \Gamma[B^0  (t)\rightarrow J/\Psi K_S] + 
      \Gamma[\overline{B^0}  (t)\rightarrow J/\Psi K_S] }\\
 &=&  \frac{2}{(2+c_d)} \left[\mbox{Im}(\frac{q}{p} \rho) \sin (\Delta M_B t)
                              -\frac{c_d}{2}\cos (\Delta M_B t)\right] ,
\end{eqnarray}
where 
\begin{eqnarray}
  \frac{q}{p} \equiv \frac{|M_{12}^B|}{M_{12}^B},&\quad&
  M_{12}^B \equiv \langle B^0|{\cal H}^{\Delta B=2}| \overline {B^0}\rangle, \\
  \rho  \equiv  \frac{A(\overline{B^0}\rightarrow J/\Psi K_S)}{
                        A(B^0\rightarrow J/\Psi K_S)},&\quad& 
  |\rho|^2 \equiv 1 + c_d, 
\end{eqnarray}
and we have neglected the absorptive part of $\langle B^0|{\cal H}^{\Delta B=2}| 
\overline {B^0}\rangle$, which is a good approximation in $B$ meson system. 
The assumption given before allows us to express  $\rho$ by KM matrix elements.
\beq 
\mbox{Im}(\frac{q}{p} \rho) = \mbox{Im}\left[ 
           \frac{|M_{12}^B|}{M_{12}^B} \frac{V_{cb}{V_{cs}}^*}{{V_{cb}}^*V_{cs}}
                 \left(\frac{q_K}{p_K}\right)^* 
        \right],
\eeq
where 
\beq
\frac{q_K}{p_K} = \frac{[(M_{12}^K -(i/2)\Gamma_{12}^K)
                            (M_{12}^{K*} -(i/2)\Gamma_{12}^{K*})]^{1/2}
                            }{M_{12}^K -(i/2)\Gamma_{12}^K},
\eeq
with $M_{12}^K -(i/2)\Gamma_{12}^K \equiv \langle K^0|{\cal H}^{\Delta S=2}|
 \overline {K^0}\rangle$. It is experimentally known that $CP$ violation in 
$K$ meson system is tiny, $O(10^{-3})$, so that we neglect it here. 
Then we can take $ M_{12}^K/\Gamma_{12}^K$ to be real, and 
\beq
\frac{q_K}{p_K} = \frac{|\Gamma_{12}^K|}{\Gamma_{12}^K}
               =  \frac{V_{ud}{V_{us}}^*}{{V_{ud}}^*V_{us}}.
\eeq
The phase of $\Gamma_{12}^K$ is calculated from  $W$ boson exchange 
tree decay diagram since we neglected $CP$ violation in $K$ meson system. 
Let us define the phase discrepancy between the KM factors as  
\beq
\delta_1 \equiv \mbox{arg}[V_{ud}{V_{us}}^*] -\mbox{arg}[V_{cd}{V_{cs}}^*] + \pi ,
\eeq
where $\delta_1=O(10^{-3})$ in the 3 generation standard model.
We have
\beq
 \frac{Asy[J/\Psi K_s]}{\sin (\Delta M_B t)} = 
 \mbox{Im}\left[  \frac{|M_{12}^B|}{M_{12}^B}
       \frac{V_{cb}{V_{cs}}^*}{{V_{cb}}^*V_{cs}}
       \frac{{V_{cd}}^* V_{cs}}{V_{cd}{V_{cs}}^*} e^{-2i\delta_1}\right]
        = -\sin (\phi_M + 2 \phi_c + 2 \delta_1), \label{phi1}
\eeq
where $\phi_M\equiv \mbox{arg}[M_{12}^B]$, $\phi_c \equiv \mbox{arg}[{V_{cb}}^* V_{cd}]$.
In the case of the 3 generation standard 
model $\phi_M = - 2  \mbox{arg}[{V_{tb}}^* V_{td}]$, so that 
the righthand-side of eq.(\ref{phi1}) becomes $-\sin 2 \phi_1$ up to tiny 
correction of $\delta_1$. If a new physics affects $\phi_M$ or $\delta_1$, then 
the $CP$ angle to measure, 
\beq
\tilde \phi_1 =\frac{1}{2}\phi_M +\phi_c +\delta_1-\pi \mbox{ (mod $\pi$)}, 
\eeq
can deviate from the geometrical angle, $\phi_1$. This CP angle $\tilde \phi_1$ is 
rephasing invariant because $M_{12}^B \rightarrow e^{2i(\hat b-\hat d)}M_{12}^B$ under 
the redefinition of quark phases.

\subsection{$\phi_2 \ (\alpha)$}
The $CP$ angle $\tilde\phi_2$ is measured in a similar manner as  $\tilde\phi_1$ 
by using $b(\bar b) \rightarrow u\bar u d (\bar d)$ decay. A typical CP 
eigenstate is $\pi\pi$. There is a $\Delta I =1/2$ penguin 
contribution in this decay mode. But it can be removed by isospin 
analysis\cite{iso}. The weak phase of the resulting decay amplitude is 
controlled by KM matrix elements following the assumption. 
The $CP$ asymmetry is given by 
\beq 
\left.
\mbox{Im}(\frac{q}{p} \rho)\right|_{\pi\pi} = -\mbox{Im}\left[ 
           \frac{|M_{12}^B|}{M_{12}^B} \frac{V_{ub}{V_{ud}}^*}{{V_{ub}}^*V_{ud}}
        \right] = \sin (\phi_M +2\phi_u),
\label{pipi}
\eeq
where  $\phi_u \equiv \mbox{arg}[{V_{ub}}^* V_{ud}]$.
In the case of the standard model the righthand-side of eq.(\ref{pipi}) becomes 
$\sin [2(\pi- \phi_2)]=- \sin 2\phi_2$. With new physics effects on  $M_{12}^B $, 
the $CP$ angle to measure, 
\beq
\tilde\phi_2 = -\frac{1}{2}\phi_M -\phi_u +\pi \mbox{ (mod $\pi$)}, 
\eeq
can deviate from $\phi_2$. This $\tilde\phi_2$ is also rephasing invariant 
as it should be.

\subsection{$\phi_3 \ (\gamma)$}
The rest of the $CP$ angles $\tilde\phi_3$ is to be measured at B factories
from the decays 
$B^{\pm} \rightarrow \{D^0, \overline{D^0}, D_{CP}\}K^{\pm}$ or 
$B^0 ( \overline{B^0})\rightarrow \{D^0, \overline{D^0}, D_{CP}\}K_S$ \cite{phit}, 
where $D_{CP}$ is a CP eigenstate of neutral $D$ meson. 
The CP eigenstate $D_{CP}$ is given generally as
\beq
D_{CP} \equiv  e^{i\xi_D}( D^0 \pm e^{i\varphi_D}\overline{D^0})/\sqrt{2},
\eeq 
by defining $CP |D^0 \rangle = e^{i\varphi_D} |\overline{D^0} \rangle$.
The overall phase $e^{i\xi_D}$ is irrelevant, so omitted below. The meson 
CP phase $e^{i\varphi_D}$ can be taken arbitrary, but 
changes its value under the rephasing of quarks. To see this let us take 
\beq
|D^0 \rangle \sim (\bar c \gamma_5 u)|0 \rangle, \mbox{\ \ \ } 
|\overline D^0 \rangle \sim e^{i\chi}(\bar u \gamma_5 c)|0 \rangle, 
\label{ddef}
\eeq
where $\sim $ means that r.h.s and l.h.s have the same quantum number. 
The phase $e^{i\chi}$ can be taken arbitrary. Then we have
\beq
CP|D^0 \rangle \sim -e^{i(\tilde u - \tilde c)} (\bar u \gamma_5 c)|0 \rangle
               \sim e^{i(\tilde u - \tilde c-\chi +\pi)} |\overline D^0 \rangle.
\eeq
So the meson CP phase given by 
$\varphi_D = \tilde u - \tilde c-\chi +\pi$ has another 
arbitrary factor $\chi$ other than quark CP phases. It varies  under rephasing
as $\varphi_D' =  \varphi_D +2(\hat c -\hat u)$ as quark CP phases transform.

The decay amplitudes of $B^+ \rightarrow D^0K^+$ and 
$B^+ \rightarrow \overline{D^0}K^+$ are written as
\beqar
A(B^+ \rightarrow D^0K^+) &=& {V_{ub}}^*V_{cs}a_D e^{i\delta_D},\\
A(B^+ \rightarrow \overline{D^0}K^+) &=& {V_{cb}}^*V_{us}a_{\bar D}
                         e^{i\delta_{\bar D}},
\eeqar
where the KM matrix elements and strong phase shifts ($\delta_D$, $\delta_{\bar D}$) are 
factored out. Note that the rest of the amplitudes ($a_D$, $a_{\bar D}$) are not 
necessarily real depending on the convention (\ref{ddef}). 
\beq
e^{-i\varphi_D}\frac{a_{\bar D}}{a_D} = - e^{-i(\tilde u -\tilde c)}
\times \mbox{(real factor)}.
\eeq
The partial decay 
width of $B^+$ decay into $K^+$ and CP even eigenstate of $D$ meson is given as 
\beqar
\Gamma( B^+ \rightarrow  D_{CP+}K^+) &=&
\frac{1}{2}
[\Gamma( B^+ \rightarrow  D^0K^+)+ \Gamma( B^+ \rightarrow  {\overline D^0}K^+)]
\nonumber \\
& & 
+\int d\Gamma |V_{ub}V_{cs}|^2
\mbox{Re}\left[
\frac{{V_{cb}}^* V_{us}}{{V_{ub}}^* V_{cs}} a_D^*a_{\bar D} e^{-i(\varphi_D +\delta_s)}
\right],
\eeqar
where $\delta_s = \delta_D - \delta_{\bar D}$. 
The final term depends on 
\beq
\arg \left[\frac{{V_{cb}}^* V_{us}}{{V_{ub}}^* V_{cs}}
             e^{-i(\tilde u -\tilde c - \delta_s)}
      \right],
\label{cpdk}
\eeq
which is rephasing invariant. Only $V_{ub}$ and $e^{i\delta_s}$ 
have non-negligible phases in the 3 generation standard model 
with vanishing quark CP phase convention ($\tilde q =0$) and the Wolfenstein parameterization  
of KM matrix\cite{wolf}, so we can get the angle $\phi_3$ by combining 
this with CP conjugate decays. 
If the condition (\ref{cpkm}) is satisfied, the weak phase vanishes 
in the above expression (\ref{cpdk}) and we have no CP violation. 
But it seems strange that the physical 
measure of CP violation (\ref{cpdk}) depends on the arbitrary quark CP phases, 
$\tilde u$ and $\tilde c$, in the general phase convention. 
To solve this paradox let us remember how we identify the D meson CP 
eigenstate in experiment. The CP even eigenstate can be identified by its 
decay into a $K$ meson pair or a $\pi$ meson pair. Let us consider the amplitude 
of CP odd eigenstate of D meson decaying into a K meson pair;
\beq
A( D_{CP-} \rightarrow K^+K^-)
 \propto  \left[
   \frac{{V_{cs}}^*V_{us}}{V_{cs}{V_{us}}^*} e^{i(\tilde c - \tilde u)} -1 
     \right],
\label{dkk}
\eeq
because
\beqar
\langle K^+K^- |{\cal H}|D^0 \rangle &=& \langle K^+K^- |(CP)^{-1}(CP){\cal H}(CP)^{-1}(CP)
            |D^0 \rangle  \nonumber \\ 
&=& \langle K^+K^- |{\cal H}^{CP}|\overline{D^0} \rangle e^{i\varphi_D}=
\frac{{V_{cs}}^*V_{us}}{V_{cs}{V_{us}}^*} e^{i\varphi_D}e^{i(\tilde c - \tilde u)} 
\langle K^+K^- |{\cal H}|\overline{D^0} \rangle ,
\eeqar
for ${\cal H} \propto {V_{cs}}^*V_{us} \overline{s_L}\gamma_\mu c_L 
 \overline{u_L}\gamma^\mu s_L +\mbox{(h.c.)}$. 
The amplitude (\ref{dkk}) should vanish under our assumption of negligible CP 
violation in D decay, then we have 
\beq
  e^{i(\tilde c - \tilde u)} =
\frac{V_{cs}{V_{us}}^*}{{V_{cs}}^*V_{us}.}  
\eeq
Now the weak phase in the formula (\ref{cpdk}) is given as
\beqar
-\arg \left[
    \frac{{V_{cb}}^* V_{us}}{{V_{ub}}^* V_{cs}}
    \frac{V_{cs}{V_{us}}^*}{{V_{cs}}^*V_{us}}
    \right]
&=& -\arg \left[
{(V_{cb}}^*V_{cd})( V_{ub} {V_{ud}}^*)( {V_{us}}^* V_{ud})( {V_{cs}} {V_{cd}}^*)
\right] \nonumber \\
&=& \phi_u - \phi_c -\delta_1 +\pi, 
\eeqar
which becomes the CP angle $\tilde\phi_3$ to measure. 
When a $\pi$ meson pair is used to identify $D_{CP}$, the results becomes as follows
\beqar
-\arg \left[
    \frac{{V_{cb}}^* V_{us}}{{V_{ub}}^* V_{cs}}
    \frac{V_{cd}{V_{ud}}^*}{{V_{cd}}^*V_{ud}}
    \right]
&=& -\arg \left[
{(V_{cb}}^*V_{cd})( V_{ub} {V_{ud}}^*)( V_{us} {V_{ud}}^*)( {V_{cs}}^* {V_{cd}})
\right] \nonumber \\
&=& \phi_c - \phi_u +\delta_1 -\pi .
\eeqar
These two results differ by $2\delta_1$ (mod $2\pi$). But as far as our assumption holds,
it should be negligible.

\section{Concluding remarks}
We have discussed CP invariance condition and rephasing invariance in KM scheme 
of CP violation in this note without assuming the 3 generation standard model.
The CP violation measure in B physics, corresponding to the three angles of unitarity 
triangle are given in rephasing invariant fashion as follows;
\beqar
\tilde \phi_1 &=& \frac{1}{2}\phi_M +\phi_c +\delta_1-\pi,\\
\tilde \phi_2 &=& -\frac{1}{2}\phi_M -\phi_u +\pi,\\
\tilde \phi_3 &=& \phi_u - \phi_c -\delta_1 +\pi .
\eeqar
where $\phi_x =\arg [{V_{xb}}^*V_{xd}]$, $\phi_M =\arg [
 \langle B^0|{\cal H}^{\Delta B=2}| \overline {B^0}\rangle]$, 
$\delta_1 =  \arg[V_{ud}{V_{us}}^*] -\arg [V_{cd}{V_{cs}}^*] + \pi$, 
and $D^0, \overline{D^0} \rightarrow K$ meson pair is used in 
obtaining $\tilde\phi_3$. The sum of these three angels becomes $\pi$ 
though we have not assumed 3 generation. Therefore, we should use the 
informations of sides also to explore a new physics. One more interesting 
point is that the measure $\tilde \phi_3$ can differ depending on what mode 
is used to identify neutral D meson. The difference might be observable 
in a future experiment with a new physics which can contribute 
significantly CP violation in D meson system.
A quantitative discussion of this point will be given elsewhere.\cite{kt}. 


\newpage
\begin{thebibliography}{99}
\bibitem{KM} M.~Kobayashi and T.~Maskawa, Prog.Theor.Phys. {\bf 49},
652 (1973).
\bibitem{unit} For a short review see the article by H.~Quinn in
{\it Review of Particle Physics}, Patricle Data Group, Phys. Rev.  
{\bf D54}, 1 (1996).
\bibitem{new} For reviews see \\
             Y.~Nir, {\it Proc. of the Workshop on B physics at Hadron Accelerators}, 
             eds. P.~McBride and C.S.~Mishra, SSCL-SR-1225, Fermilab-CONF-93/267, 
             185 (1993);\\
            Y.~Nir and H.R.~Quinn, Ann. Rev. Nucl. Part. Sci. {\bf 42} 211 (1992);\\
            C.O.~Dib, I.~Dunietz, F.J.~Gilman and Y.~Nir, 
            Phys. Rev. {\bf D41}, 1522 (1990).
\bibitem{KEKB} {\it Letter of Intent for a Study of CP 
                Violation in B Meson Decays}, KEK Report 94-2 (1994).
\bibitem{SLACB} {\it Letter of Intent for the Study of CP 
                  Violation and Heavy Flavor Physics at PEP II}, SLAC-443 (1994).
\bibitem{bbss} J.-M.~G\'{e}rard, W.~Grimus, A.~Masiero, D.V.~Nanopoulos,
               and A.~Raychaudhuri, Nucl. Phys. {\bf B253}, 93 (1985);\\
               M.~Dugan, B.~Grinstein, and L.~Hall, Nucl. Phys. {\bf B255}, 413 (1985);\\
               G.~Altarelli and P.~Franzini, Z. Phys. {\bf C37}, 241 (1988);\\
               T.~Kurimoto, Phys. Rev. {\bf D39}, 3447 (1989);\\
               G.C.~Branco, G.C.~Cho and Y.~Kizukuri, Phys.Lett. {\bf B337}, 316 (1994);\\
               T.~Goto, T.~Nihei and Y.~Okada, Phys.Rev. {\bf D53}, 5233 (1996); 
                 errutum-ibid. {\bf D54}, 5904 (1996).
\bibitem{lrmd}  D.~London and D.~Wyler, Phys. Lett. {\bf B232}  503 (1989), \\
                T.~Kurimoto, A.~Tomita and S.~Wakaizumi, Phys. Lett. {\bf B381} 470 (1996).
\bibitem{fgen} T.~Hasuike, T.~Hattori, T.~Hayashi, S.~Wakaizumi, 
                Mod.Phys.Lett. {\bf A4}, 2465 (1989).
\bibitem{vecq} G.C.~Branco, P.A.~ Parada, T.~Morozumi, M.N.~, Rebelo, 
               Nucl. Phys. Proc. Suppl. {\bf 37A}, 29 (1994).
\bibitem{cpang} A.B.~Carter and A.I.~Sanda, Phys. Rev. Lett. {\bf 45} 952 (1980),
               Phys. Rev. {\bf D23} 1567 (1981); \\
               I.I.~Bigi and A.I.~Sanda, Nucl. Phys. {\bf B193} 85 (1981), 
               Nucl. Phys. {\bf B281} 41 (1987).
\bibitem{iso} M.~Gronau and D.~London, Phys. Rev. Lett. {\bf 65} 3381 (1990);\\
              H.J.~Lipkin, Y.~Nir, H.R.~Quinn and A.E.~Snyder, Phys. Rev. {\bf D44} 
              1454 (1991).
\bibitem{phit} M.~Gronau and D.~London, Phys. Lett. {\bf B253} 483 (1991);\\
               M.~Gronau and D.~Wyler,  Phys. Lett. {\bf B265} 172 (1991);\\
               I.~Dunietz, Phys. Lett. {\bf B270} 75 (1991).
\bibitem{wolf} L.~Wolfenstein, Phys. Rev. Lett. {\bf 51} 1945 (1983).
\bibitem{kt} T.~Kurimoto and A.~Tomita, in preparation.
\end{thebibliography}
\end{document}